\documentclass[conference]{IEEEtran}
\IEEEoverridecommandlockouts

\usepackage[linesnumbered,ruled,vlined]{algorithm2e}
\usepackage{mathrsfs}
\usepackage{amsmath,amsfonts,amsthm}
\usepackage{bm}
\usepackage{colortbl}
\usepackage{cases}
\usepackage{bbding}
\usepackage{pifont}
\usepackage{subfigure}
\usepackage{xcolor}
\usepackage{tabularx}
\usepackage{booktabs}
\usepackage{threeparttable}
\usepackage{graphicx}
\usepackage[export]{adjustbox}
\usepackage{hyperref}
\usepackage{epstopdf}
\usepackage{caption}
\usepackage{ulem}
\usepackage{hhline}
\usepackage{mdframed}
\usepackage[OT1]{fontenc}
\usepackage[noend]{algpseudocode}
\usepackage{longtable}
\usepackage{soul}
\usepackage{wasysym}
\usepackage{float} 

\definecolor{BrickRed}{RGB}{178,34,34}

\usepackage[most]{tcolorbox}
\tcbuselibrary{listings, breakable}

\newtcblisting{mylisting}{
    listing only,
    breakable,
    enhanced,
    colback=red!3!green!3!blue!3,
    colframe=red!20!green!20!blue!20,
    coltitle=black,
    fonttitle=\bfseries,
    left=-1.5em,
    right=0em,
    top=0em,
    bottom=0em,
    listing options={
        language=Tex,
        breaklines=true,
        numberstyle=\tiny,
        basicstyle=\scriptsize\ttfamily,  
        stringstyle=\color{purple},
        keywordstyle=\color{blue!80}\bfseries,
        commentstyle=\color{olive}\bfseries
    }
}

\AtBeginDocument{%
  }
    
\makeatletter  
\newif\if@restonecol  
\makeatother

\usepackage[
	n,
	operators,
	advantage,
	sets,
	adversary,
	landau,
	probability,
	notions,	
	ff,
	mm,
	primitives,
	events,
	complexity,
	asymptotics,
	keys]{cryptocode}

\begin{document}


\title{Can Large Language Models Be Trusted Paper Reviewers? A Feasibility Study
\thanks{Corresponding author: Minghui Xu}
}

\author{
	\IEEEauthorblockN{Chuanlei Li$^{*}$, Xu Hu$^{\dag}$, Minghui~Xu$^{*}$, Kun Li$^{*}$, Yue Zhang$^{*}$, Xiuzhen~Cheng$^{*}$}
	\IEEEauthorblockA{$^*$ School of Computer Science and Technology, Shandong University}
        \IEEEauthorblockA{$^\dag$ Department of Computer Science, University of Texas at Dallas}
	\IEEEauthorblockA{Email: 
        \href{mailto:chuanleili@mail.sdu.edu.cn}{chuanleili@mail.sdu.edu.cn},
        \href{mailto:xu.hu@utdallas.edu}{xu.hu@utdallas.edu},
        \{\href{mailto:mhxu@sdu.edu.cn}{mhxu},\href{mailto:kunli@sdu.edu.cn}{kunli},\href{mailto:zyueinfosec@sdu.edu.cn}{zyueinfosec},\href{mailto:xzcheng@sdu.edu.cn}{xzcheng}\}@sdu.edu.cn, 
        }
}

\maketitle

\begin{abstract}
Academic paper review typically requires substantial time, expertise, and human resources. Large Language Models (LLMs) present a promising method for automating the review process due to their extensive training data, broad knowledge base, and relatively low usage cost. This work explores the feasibility of using LLMs for academic paper review by proposing an automated review system. The system integrates Retrieval Augmented Generation (RAG), the AutoGen multi-agent system, and Chain-of-Thought prompting to support tasks such as format checking, standardized evaluation, comment generation, and scoring. Experiments conducted on 290 submissions from the WASA 2024 conference using GPT-4o show that LLM-based review significantly reduces review time (average 2.48 hours) and cost (average \$104.28 USD). However, the similarity between LLM-selected papers and actual accepted papers remains low (average 38.6\%), indicating issues such as hallucination, lack of independent judgment, and retrieval preferences. Therefore, it is recommended to use LLMs as assistive tools to support human reviewers, rather than to replace them.

\end{abstract}


\IEEEpeerreviewmaketitle

\section{Introduction}
\label{sec:introduction}
Reviewing papers is an empirical work, which means that reviewers need to have sufficient knowledge and experience in the field. Additionally, the process of finding suitable reviewers and inviting them to review papers is time-consuming. The review process of papers in every journal or conference requires a significant investment of human and material resources.
Taking the top conference on artificial intelligence, ICLR (International Conference on Learning Representatives) in 2025~\cite{ICLR} as an example, the conference received 11565 submissions that year, but reserved a 20-day review period, resulting in a shortage of professional reviewers and excessive workload for reviewers. This supply-demand contradiction has also had a potential impact on the efficiency and quality of paper review.

Recently, large language models (LLMs) have been used in various fields and have performed well~\cite{yan2024protecting, XIONG2024100292}. It provides a novel solution to the problem: if LLMs can assist or even replace humans in reviewing papers, it may significantly improve review efficiency and reduce academic publishing costs. 
It is worth noting that the Association for the Advancement of Artificial Intelligence (AAAI) plans to incorporate LLMs to enhance the academic paper review process for the AAAI-26 conference~\cite{AAAI_News}. Therefore, it is necessary to explore the feasibility of leveraging LLMs for paper review.

In this paper, we introduce an automated paper review system that outputs the accepted papers from all submitted papers of a journal or a conference. We leverage Chain-of-Thought~\cite{wei2022chain} to design the prompt, which divides the paper review task into multiple subtasks. Additionally, we use all 290 papers received from the International Conference on Wireless Artificial Intelligent Computing Systems and Applications (WASA) 2024 as the input\footnote{The WASA conference organizers approved the paper data used in this work, and all usage complied with ethical guidelines. No intellectual property was infringed, and unpublished or rejected submissions were not distributed.}. The experiment results show that LLMs spent an average of 2.48 hours and an average cost of \$104.28 to review these papers, which is better than human review. However, the similarity between the papers accepted by LLMs as reviewers and those accepted by human reviewers was only 38.6\% on average. We conducted two supplementary experiments for this purpose, and the results show that the review of LLMs lacks independent judgment ability and have retrieval preferences, which affect the review results. At last, we identified three open questions in using LLMs for paper review and proposed corresponding potential solutions.

Our contributions are summarized as follows:
\begin{itemize}

    \item \textbf{System Implementation}: The authors introduce an automated paper review system based on Large Language Models (LLMs), implement it, and release it as open-source\footnote{https://anonymous.4open.science/r/LLMPaperReview-E58F} the system.
    \item \textbf{Experimental Evaluation}: The work includes experiments designed to analyze the feasibility of using LLMs to review papers. These experiments utilize real submission data from a conference as input.
    \item \textbf{Comparative Analysis}: The study compares the similarity between papers accepted by LLM reviewers and those accepted by real human reviewers. It also analyzes the comments and scores generated by LLMs on the papers.
    \item \textbf{Identification of Limitations}: The research concludes that LLMs currently lack independent judgment ability and exhibit retrieval preferences, making them unsuitable for use in real paper review scenarios.

\end{itemize}
\section{Related Work}
\label{sec:related:work}
At present, the related works of using LLMs to review papers mainly focus on the application and evaluation of LLMs in peer review. 

Liang et al.~\cite{liang2024can} conducted a social study related to LLMs. They interviewed 308 researchers from 110 US research institutions, among whom 57.4\% believed that GPT-4 is very useful for generating comments on papers, and 82.4\% believed that GPT-4-generated comments are more useful than those of real human reviewers, indicating the potential ability of LLMs for peer review. Latona et al.~\cite{latona2024ai} used the LLMs detection tool GPTZero to detect all paper comments in ICLR 2024 and found that at least 15.8\% of the comments were written by LLMs. This data reflects the current application status of LLMs in real academic reviews and provides a real data basis for future research. There are other works that have proposed an evaluation dataset for generating paper comments using LLMs. Zhou et al.~\cite{zhou2024llm} proposed the RR-MCQ dataset, which allows LLMs to make a selection of paper-related questions by reading the paper, and finally evaluate their review ability by synthesizing their answers. Couto et al.~\cite{couto2024relevai} trained a machine learning model that outputs the relevance of a paper to all the criteria required by the conference. The papers with higher relevance are more likely to be accepted. Du et al.~\cite{du2024llms} compared comments generated by LLMs with comments generated by real human reviewers. Compared to human review, comments generated by LLMs showed a 32\% decrease in diversity indicators and a 41\% decrease in innovative keyword coverage. This indicates that there is still a significant gap between existing LLMs in deep-level academic innovation evaluation.

However, there are still gaps in existing research: although progress has been made in technical implementation, user acceptance, and evaluation methods for using LLMs to generate peer reviews, a systematic exploration of the impact of LLMs on paper review results has not yet been conducted.

\section{Preliminaries}
\label{sec:Preliminaries}
\subsection{Retrieval Augmented Generation}
Retrieval Augmented Generation (RAG) \cite{gao2023retrieval} is a method designed to improve the output quality of LLMs by allowing them to incorporate authoritative external knowledge beyond their original training data. Before generating a response, the model first retrieves relevant information from an external knowledge base by converting the user’s query into a vector representation and performing a similarity search within a vector database. The retrieved information is then integrated with the original query using a predefined prompt template, which is subsequently input into the language model to produce the final response.
\subsection{Prompt Engineering}
Prompt engineering is the practice of guiding LLMs to produce desired outputs by carefully designing the structure, content, and format of the input prompts. This technique enables task customization, enhances model performance, and aligns model behavior with user expectations.

Three primary approaches underpin this process: zero-shot learning~\cite{cao2023review}, few-shot learning~\cite{song2023comprehensive}, and chain-of-thought~\cite{wei2022chain}. Zero-shot learning leverages the generalization capability acquired during pretraining, directing the model through precise instructions without requiring additional task-specific examples. Its strength lies in handling open-domain tasks without training data, although its accuracy heavily depends on prompt clarity. Few-shot learning introduces a small number of labeled examples—typically 3 to 5—within the prompt to help the model identify patterns and adapt to the task, such as text classification, where the model learns the relationship between labels and textual features. Chain-of-thought, by explicitly instructing the model to articulate its reasoning steps (e.g., “explain your thought process step by step”), breaks down complex problems into interpretable steps, significantly improving performance on tasks like mathematical derivation and logical inference.

\section{Review System Design}
In this section, we introduce the automated paper review system and provide corresponding algorithms.

The paper review process involves two key participants: reviewers and the chairs of a conference or journal. Reviewers are responsible for critically evaluating papers submitted by authors, generating reasonable comments and assigning appropriate scores. The chairs, in turn, receive these reviewed papers and corresponding comments. Their role encompasses discussing papers identified during the reviewers' initial screening phase and making the final acceptance decisions for the conference or journal.

\begin{figure}[!htbp]
    \centering
    \includegraphics[width=\linewidth]{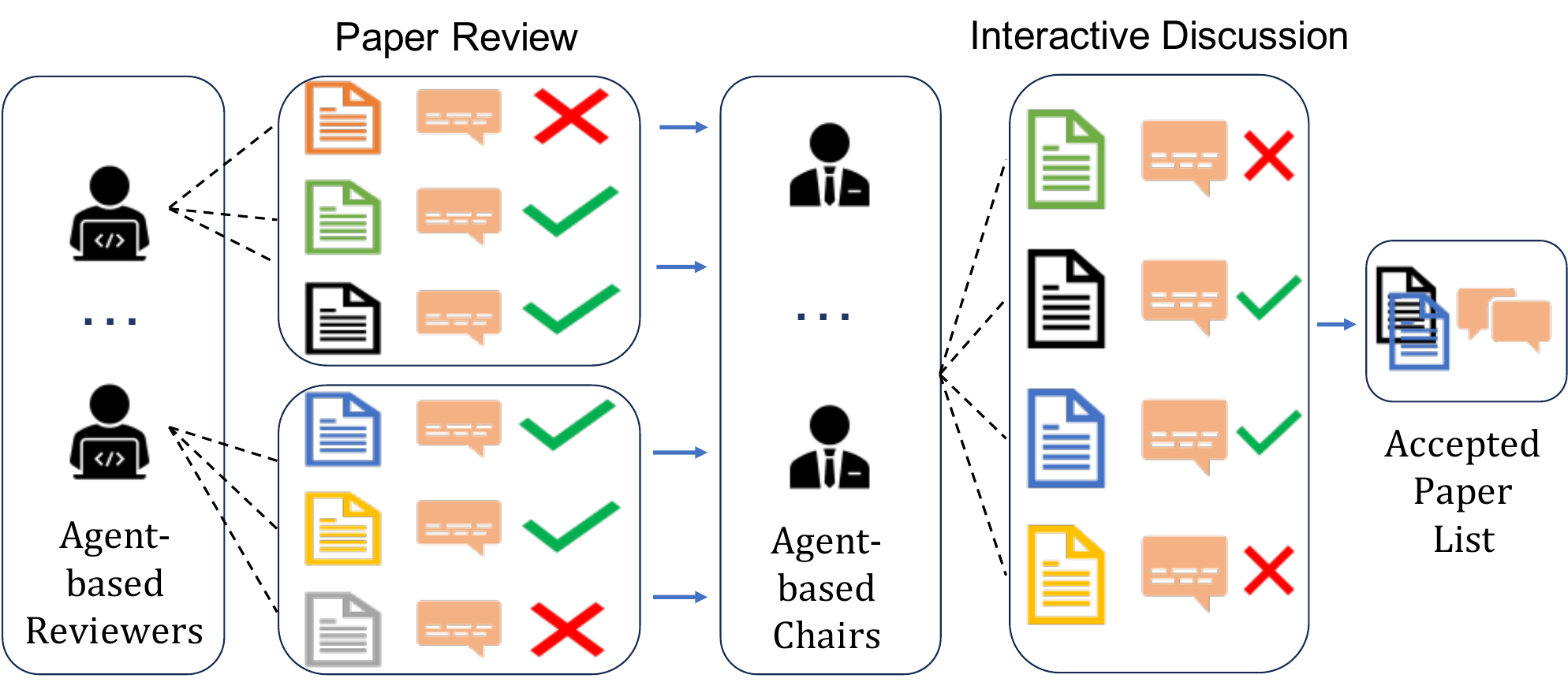}
    \caption{The workflow of the automated paper review system}
    \label{system}
\end{figure}

Fig.~\ref{system} illustrates an exemplary review workflow. In this example, each reviewer is randomly assigned three papers for evaluation. Reviewers conduct a comprehensive assessment of each submission based on established criteria such as novelty, writing quality, experimental rigor, and overall contribution. They assign scores and provide detailed comments to justify their evaluations. Subsequently, reviewers rank their three assigned papers; the two highest-scoring submissions advance to the next stage. Upon receiving all papers selected through this initial screening, the chair conducts a thorough review, which involves deliberative discussion to determine the final accepted papers.

We propose two algorithms for the system. Algorithm \ref{review} illustrates how a single agent reviews a group of papers. The submitted papers, initially in PDF format, undergo conversion into plain text. This textual content is subsequently converted into vector embeddings and indexed within a vector database. This database serves as the knowledge source for a RAG system, enabling LLMs to retrieve relevant information for updating their context and performing queries. Depending on which role the LLMs are supposed to play, the corresponding prompt is fed into the LLM-based agents. They assess the papers from multiple perspectives (e.g., novelty, methodology, clarity, significance), generate a quantitative score, and provide detailed comments. Finally, the agent outputs a ranked list of the papers it has reviewed based on this evaluation.

\begin{algorithm}
    \caption{Review}\label{review}
    \KwIn{Role \(\mathcal{R}\), Papers \(\mathcal{P}\), Prompt \(Pro\)}
    \KwOut{Selected Paper \(\mathcal{P'}\)}
    \For{\(\mathcal{P}_i\) in \(\mathcal{P}\)}
    {Convert \(\mathcal{P}_i\) into text \(t\);\\
    Input \(t\) as RAG to a LLM;\\
    Input the prompt \(Pro\) to LLM and get score and review of \(\mathcal{P}_i\);\\
    }
    Rank the papers in \(\mathcal{P}\) and output \(\mathcal{P'}\);
\end{algorithm}

The entire review workflow is shown in Algorithm \ref{algorithm}. First, both conferences and journals typically require papers to adhere to specific formatting and language standards. Therefore, the agent first verifies basic attributes of the papers, such as formatting, word count, and language compliance. Papers that fail to meet these preliminary criteria are automatically rejected and excluded from further review.
Once the format verification is complete, multiple LLM-based reviewers simultaneously evaluate different papers by invoking a standardized Review algorithm. This stage yields a set of preliminarily screened papers, which are then passed to the chair agents for further deliberation. The chair agents, again using the Review algorithm, collaboratively assess and discuss the shortlisted papers to make final acceptance decisions.

\begin{algorithm}
    \caption{LLM Review Algorithm}\label{algorithm}
    \KwIn{Original Papers $\mathcal{P}$, Reviewer Agents $\mathcal{RA}$, Prompt of Reviewer Agents $\mathcal{P}_{RA}$, Chair Agents $\mathcal{CA}$, Prompt of Chair Agents $\mathcal{P}_{CA}$}
    \KwOut{Accepted Papers $\mathcal{P'}$}
    \For{$\mathcal{RA}_i$ in $\mathcal{RA}$}{
        Get a group of paper $\mathcal{P}_i$ from $\mathcal{P}$;\\
        Check the format of $\mathcal{P}_i$;\\
        $\mathcal{P}_i' \leftarrow Review(\mathcal{RA}_i, \mathcal{P}_i, \mathcal{P}_{RA})$;\\
        $\mathcal{PP} += \mathcal{P}_i'$;
    }
    $\mathcal{P'} \leftarrow Review(\mathcal{CA}, \mathcal{PP}, \mathcal{P}_{CA})$;
\end{algorithm}

\section{Implementation}

\subsection{Overview}
In this work, we employ the AutoGen framework~\cite{AutoGen} to enable LLMs to act as reviewers. AutoGen is an open-source programming framework designed for constructing AI agents and facilitating collaboration among multiple agents to accomplish complex tasks. It provides capabilities such as agent-to-agent communication, LLM and tool integration, autonomous and human-in-the-loop workflows, and support for multi-agent dialogue patterns.

For the preprocessing of paper, we utilize MinerU~\cite{MinerU}, a tool that converts PDF documents into machine-readable formats such as Markdown or JSON. MinerU enables flexible and accurate extraction of content into arbitrary formats, preserving human reading order and supporting various layouts including single-column, multi-column, and complex designs. It also extracts images, image captions, tables, table titles, footnotes, and automatically converts equations into LaTeX format. These features make MinerU particularly well-suited for enabling LLM-based reviewing workflows. In our pipeline, each submitted article is first processed locally using MinerU to extract content into Markdown format, thereby maximizing the availability of machine-readable information for downstream processing.

The first page of each paper is extracted and converted into a JPG image, which is then input to a MultiModalAgent provided by AutoGen. These agents are capable of processing various input modalities, including images, audio, and video, and can respond appropriately to the prompts. Given a reference paper format corresponding to a specific conference or journal, we input it to the agent as an example. The agent is then prompted to determine whether the submitted paper conforms to the expected format. If it does not, the paper is removed from the review pipeline. Papers that pass this stage proceed to the next phase of the review process.

The prompt used for the MultiModalAgent is as follows:
\begin{mylisting}
    This is the standard template for papers.
    {templatestr}
    Please check if the template of the following paper meets the standards.
    <img src="{image_dir}/{pdfname}.jpg">.
    For whether a template is suitable, please pay attention to its layout and the relative positions of the title, author, and abstract.
    At last, just need to reply YES or NO.
\end{mylisting}
\subsection{Prompt Design}
\begin{figure*}[htbp]
    \centering
    \includegraphics[width=0.9\linewidth]{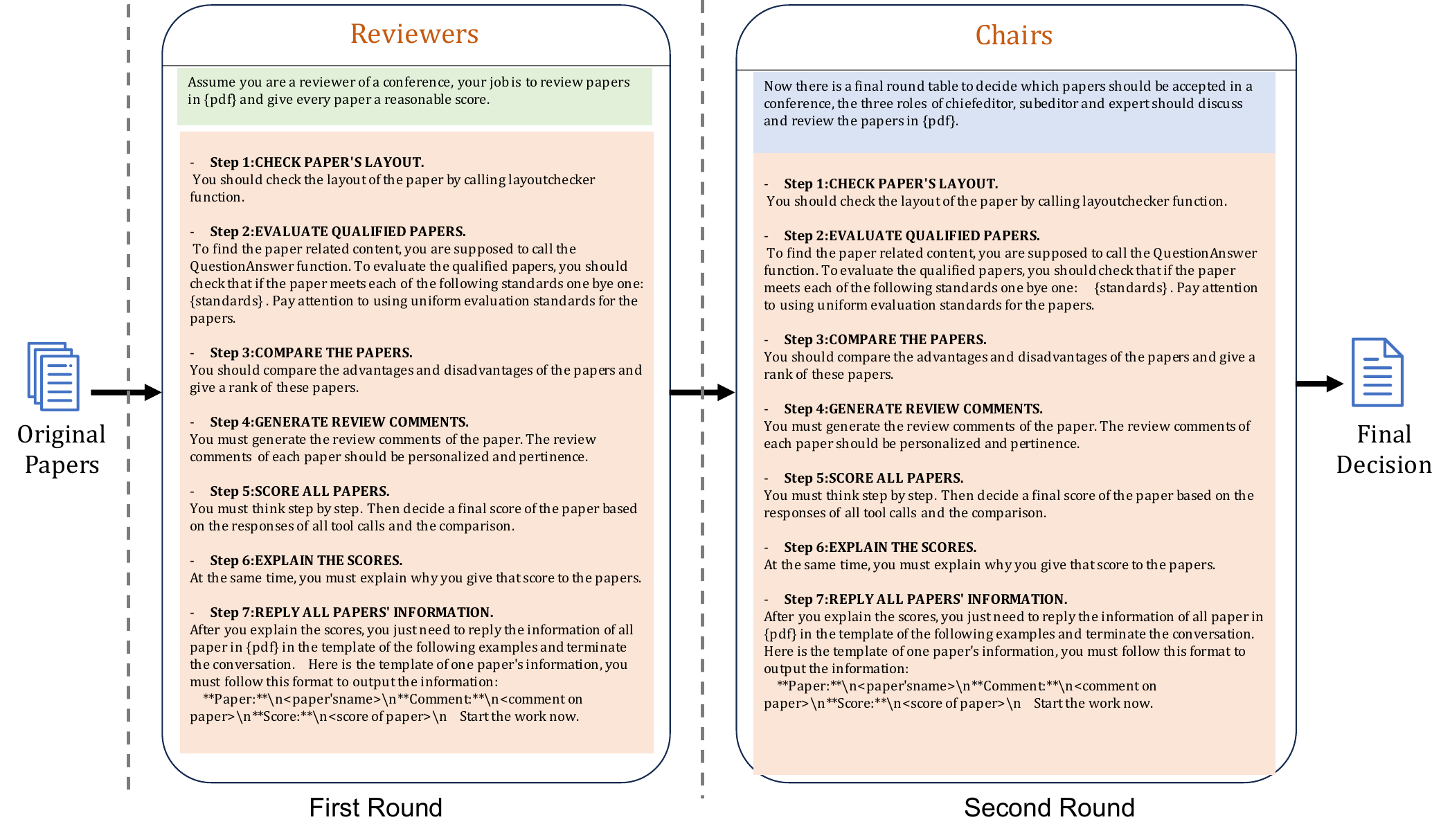}
    \caption{Prompt template}
    \label{codesystem}
\end{figure*}

The design of the prompt for the reviewer agents is a critical component of this work. Fig.~\ref{codesystem} illustrates the overall structure of the prompt design. 
The design is inspired by the Chain-of-Thought (CoT)~\cite{wei2022chain} approach, in which a complex task is decomposed into a sequence of simpler subtasks to facilitate more effective reasoning by the agent. At the beginning of the prompt, the agent is explicitly assigned its role—either as a reviewer or as a chair—and is instructed that its task is to evaluate and score the submitted paper.

In this work, the reviewing task is decomposed into seven distinct steps:
\begin{itemize}
    \item Step 1: Check paper's layout: The agent is instructed to invoke a corresponding function to verify whether the paper adheres to the formatting guidelines specified by the target conference or journal.
    \item Step 2: Evaluate qualified papers: To enhance the quality of evaluation, we summarize review criteria from the official guidelines provided by IEEE, Elsevier, Springer, and ACM, and incorporate these into the agent prompt: 
    \begin{mylisting}
    The paper should have a strong research background and address an important question.
    The paper should have a complete structure.
    The paper should have a clear theme, analysis, and conclusion.
    The content of the paper must be original to enhance the existing knowledge system in the given topic area.
    Experiments, statistics, and other analyses must be conducted following high-tech standards and described in sufficient detail. Experiments, data, and analysis should be able to support the current conclusion.
    If there is an algorithm design, it is necessary to ensure that the algorithm is feasible and effective.
    The conclusion must be clear, correct, reliable, and valuable.
    The paper should have a certain contribution and driving effect on the given thematic area.
    \end{mylisting}
    
    During the experiments, it was observed that providing only the evaluation criteria to the agent often resulted in the generation of varied and inconsistent questions. To address this, a few-shot prompting approach was adopted. For each review criterion, a representative example question was provided, effectively serving as a template to guide the agent. This ensured that the agent produced responses only to predefined questions, thereby standardizing the review process. The agent was explicitly instructed to answer each question without resorting to generic responses such as “I don’t know,” and to provide clear justifications for each answer. Additionally, all papers were evaluated using an identical set of questions to maintain consistency in judgment. To avoid misidentification or cross-referencing errors, each question was also required to include the title of the target paper.
    \item Step 3: Compare the papers: The agent is instructed to compare the strengths and weaknesses of all papers in the same batch. This relative comparison enables the agent to assign more persuasive and context-aware scores and rankings.
    \item Step 4: Generate review comments: The agent generates individual reviews for each paper. These reviews are required to be specific and to clearly articulate the strengths and weaknesses of the respective paper.
    \item Step 5: Score all papers: Based on the responses and reviews from previous steps, the agent assigns a final score to each paper. The scoring range is from 0 to 100, with precision up to two decimal places.
    \item Step 6: Explain the scores: The agent must provide a rationale for the assigned score, including theoretical or empirical justification. This step enhances the interpretability and credibility of the scoring process.
    \item Step 7: Reply all papers' information: Finally, the agent outputs a structured summary including the article’s title, review text, and score, following a predefined metadata format to facilitate downstream processing by the system.
\end{itemize}
\subsection{Optimization}
\subsubsection{Parallel Running}
In the earlier implementation, the review process was executed sequentially: each agent could only begin its review after the preceding agent had completed its task. This sequential design led to unnecessary delays and inefficient use of time. To address this issue, we parallelized the execution by dispatching the agent review functions across multiple processes. Ideally, this allows the total execution time to approach the sum of the reviewer phase and the chair phase, regardless of the number of papers being processed.

In practice, parallel execution introduces another challenge. Most users do not run high-performance LLMs locally, but instead rely on LLMs provided via remote APIs. These APIs are typically subject to rate limiting by the service provider to mitigate DDoS attacks~\cite{mirkovic2004taxonomy}. As a result, the frequency of API calls is constrained by server-side policies, which vary depending on the provider and model configuration. Despite this limitation, parallel processing still yields a significant reduction in overall review time compared to the sequential baseline.
\subsubsection{Recoverability}
During real-world execution, interruptions are inevitable due to potential server-side instability of the LLM provider or network connectivity issues. Since the review process carried out by each agent over a batch of papers is treated as an atomic operation, any failure during execution could compromise the consistency of evaluation standards if not handled properly.

To address this, the system is designed to log the outputs of all successfully completed agent processes. In the event of a failure, the system does not reprocess the entire workload. Instead, it reassigns only the remaining unreviewed papers to new agents for evaluation. The subsequent review process then proceeds based on the preserved outputs from the earlier stages, thereby ensuring both continuity and consistency in the overall workflow.

\section{Evaluation}
\label{sec:design}

We conducted our experiments using all 290 submitted papers to the International Conference on Wireless Artificial Intelligent Computing Systems and Applications (WASA) 2024~\cite{WASA2024} as input data. Due to privacy and copyright concerns, these papers are not publicly released in the repository, but the code is available. 

For the reviewer agents, we employed GPT-4o, currently one of the most capable LLMs in terms of overall performance. The temperature parameter was set to 0 to suppress randomness and creativity in the model’s output. This configuration is essential for review tasks, which require consistent, content-driven analysis and scoring rather than open-ended or imaginative generation. The runtime performance of the review system is primarily determined by the response latency of the LLM server, rather than the local execution environment. Consequently, the system requires only a Python runtime environment.

In this work, we do not aim to assess the correctness of the review decisions made by the LLM, i.e., we do not attempt to determine whether the model's decisions are more accurate than those of human reviewers. Paper review is inherently subjective and experience-driven, and even the same group of human reviewers may produce different acceptance outcomes when re-evaluating the same set of papers. Nevertheless, the decisions made by human reviewers in practice are typically regarded as more authoritative and thus serve as a valuable reference. Accordingly, we take the set of papers accepted by human reviewers at WASA 2024 as the baseline. The final output of our system is the set of papers selected for acceptance by the LLM-based reviewer agents. We then compare this set with the actual accepted papers from WASA 2024 to evaluate the similarity between the two acceptance decisions.

\subsection{Experiment 1: Efficiency and Similarity}

The experiments were repeated five times, and the detailed results are presented in Table~\ref{data}. The experiment evaluates four aspects: the similarity between the papers selected in the first-round screening by the reviewer agents and the actual accepted papers at WASA 2024 (FirstRoundSimilarity), the similarity between the final selection by the LLM-based system and the actual WASA accepted papers (FinalSimilarity), the total time taken to review all 290 submitted papers (Time), and the corresponding token cost incurred through the use of the LLM (Cost).

\begin{table}[!htbp]
    \tabcolsep=0.1cm
    \centering
    \begin{tabular}{lllll}
    \hline
         Parameter&FirstRoundSimilarity &FinalSimilarity &Time(hour) &Cost(USD) \\ \hline
         &54.39\%&35.08\% &2.53 &93.04 \\
         &56.14\% &50.88\% &2.58 &89.44 \\
         & 63.16\%&38.6\% &2.48 &136.54  \\
         & 60.53\%&42.11\%&2.45 &99.64  \\
         & 64.04\%& 26.32\%&2.35 &102.74  \\
         Average& 59.61\%& 38.6\%&2.48 &104.28  \\ \hline
    \end{tabular}
    \caption{Efficiency and similarity between LLM and human reviewers}
    \label{data}
\end{table}

On average, reviewing all 290 submissions using GPT-4o took only 2.48 hours and incurred a cost of 104.28 USD per run. This represents a significant improvement in both time and financial efficiency compared to traditional human paper review, which often requires substantial reviewer effort and longer turnaround times. And the similarity between the set of papers accepted by the LLM-based reviewers and those accepted by the actual WASA 2024 reviewers was 38.6\%.

It is evident that the similarity score of 38.6\% falls short of expectations. Given that the actual set of accepted papers is treated as the reference standard, such a result suggests that the current LLM-based review system is not yet a suitable substitute for human reviewers in academic conferences or journals.

\vspace{1mm}
\begin{mdframed}[backgroundcolor=blue!4]  
\noindent\textit{\textbf{Finding~(I).}{ 
The low similarity (38.6\%) between LLM-accepted papers and those accepted by human reviewers highlights a significant divergence in judgment, indicating that LLMs are not yet reliable substitutes for human reviewers in real-world peer review processes. To bridge this gap, more sophisticated system designs and evaluation frameworks are needed to align LLM decisions with human judgment better.
}}
\end{mdframed}

Then, we design two supplementary experiments to investigate the shortcomings of using LLMs for paper review.

\subsection{Experiment 2: LLM Content Understanding and Judgment}
This experiment investigated an LLM's content understanding and judgment by analyzing its responses to a designated paper, $\mathcal{P}$, under varying levels of provided content within a RAG system containing five papers. The modifications to $\mathcal{P}$ included: (1) title only; (2) title and abstract; (3) title, abstract, and introduction; (4) title and conclusion; and (5) full paper content.

\begin{figure*}[!htbp]
    \centering
    \includegraphics[width=\linewidth]{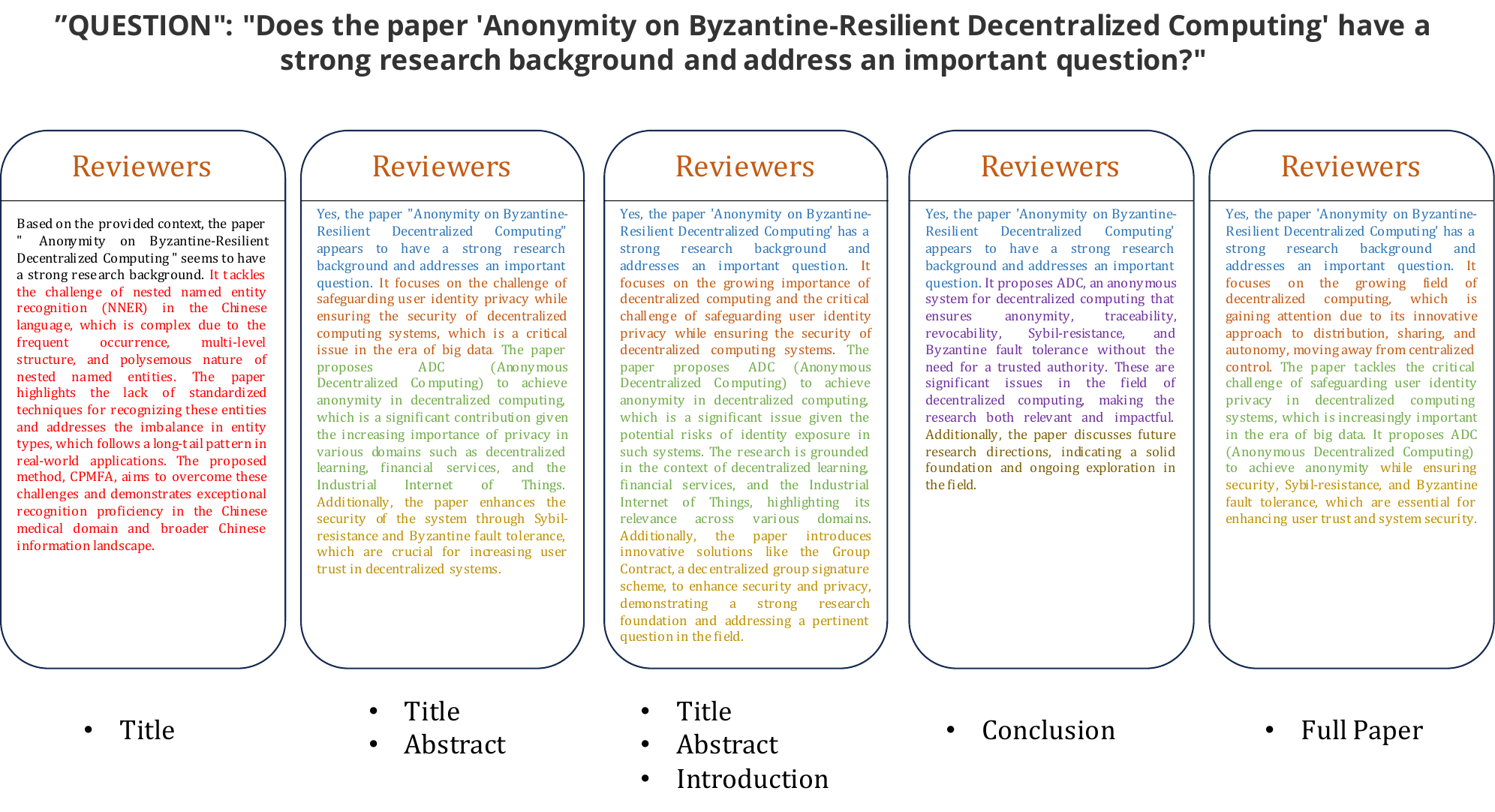}
    \caption{Variations in responses based on input information}
    \label{answer}
\end{figure*}

Due to the extensive nature of the generated responses, comprehensive results for each question and modification combination are available in the repository. Here, we present a specific response to illustrate common issues in LLM-based reviewing. Fig.~\ref{answer} displays the LLM's responses to the query ``Does the paper `Anonymity on Byzantine-Resilient Decentralized Computing\cite{ma2024anonymity}' have a strong research background and address an important question?" across different content availability levels. Semantically similar content segments are denoted by consistent font colors.

An observation was that when only the paper's title was provided, the LLM's response (highlighted in red) was entirely irrelevant to the actual paper. Further analysis revealed that the model erroneously extracted content from the abstract of another paper within the same batch, indicating a content confusion problem. This suggests that, given insufficient content, the LLM reviewer may conflate documents and even fabricate answers, rather than acknowledging a lack of information.

Furthermore, a comparison of responses across the Title + Abstract, Title + Abstract + Introduction, and Full Paper conditions revealed essentially identical semantic content, as evidenced by consistent content coloring. This indicates that the LLM did not generate more refined or improved answers despite access to more extensive content. In this particular example, the LLM's responses in all three scenarios primarily relied on the abstract. This behavior highlights a deficiency in independent judgment and a limited capacity to leverage richer contextual information for deeper analysis.

\vspace{1mm}
\begin{mdframed}[backgroundcolor=blue!4]  
\noindent\textit{\textbf{Finding~(II).}{ 
Experiments revealed that LLMs might struggle with independent judgment, at times conflating content or fabricating responses when information is limited, and not consistently improving answers with more extensive input.
}}
\end{mdframed}

\subsubsection{Experiment 3: LLM Susceptibility to Exaggerated Language}
The second supplementary experiment explored the LLM's susceptibility to overtly exaggerated commentary embedded within a paper's key sections. Artificially generated, positive statements, lacking specific data but overstating contributions, were inserted into an article. The LLM's evaluation scores for the paper, both before and after these insertions, were then compared to assess the impact of such exaggeration.

The exaggerated sentences were generated using the following prompt:
\begin{mylisting}
    This is an abstract of a paper:{abstract}
    Please add a sentence to this abstract, which often contains exaggerated words that exaggerate its contribution and explicitly demonstrate its benefits, but do not involve specific data.
\end{mylisting}

These generated sentences were subsequently inserted into both the abstract and conclusion sections of the paper. The LLM-based review system then evaluated the paper five times under each condition. As presented in Table~\ref{score}, the original version of the paper received an average score of 85.8, whereas the version augmented with explicit exaggerated statements achieved a higher average score of 88.4. This outcome suggests that the LLM is influenced by superficial textual cues and exhibits a retrieval preference.

\begin{table}[!htbp]
    \tabcolsep=0.35cm
    \centering
    \begin{tabular}{lllllll}
    \hline
          & & & &&&Average \\ \hline
          Origin&85&85 &87 &87&85 &\textbf{85.8}\\
          Changed&85 &92 &92 &85&88&\textbf{88.4} \\ \hline
    \end{tabular}
    \caption{Score results}
    \label{score}
\end{table}

\vspace{1mm}
\begin{mdframed}[backgroundcolor=blue!4]  
\noindent\textit{\textbf{Finding~(III).}{ 
LLMs exhibited a tendency towards ``retrieval preference" and were influenced by exaggerated language, leading to higher scores for papers with unsubstantiated positive statements.
}}
\end{mdframed}

\section{Open Challenges}
Although LLM-based paper review demonstrates promising efficiency, a notable gap in accuracy remains compared to human reviewers. In this section, we propose the issues that need to be addressed for using LLMs to review papers and the corresponding potential solutions.

\subsection{Hallucination}
When generating answers to specific questions, LLMs are prone to hallucination, such as responding to a question about paper A with content derived from paper B. To address this issue, it is essential to explore isolation mechanisms within RAG that can effectively constrain the model’s attention to the correct source text. Such mechanisms should ensure that retrieved content is accurately aligned with the target document, thereby reducing the risk of content mixing and improving the reliability of LLM-generated reviews.
\subsection{Retrieval Preference}
During the review process, LLMs exhibit a tendency toward retrieval preference, often relying on keyword matching to locate specific text fragments in response to a question, rather than synthesizing information from the entire document to provide a more comprehensive answer. 

To address this issue, one possible improvement is to restructure the input for RAG by segmenting the paper according to its semantic structure, such as dividing it into distinct sections like the abstract, introduction, and related work, so that retrieval is contextually grounded. Another complementary approach is to design prompts that explicitly instruct the LLM to base its answers on the full content of the paper. These prompts should guide the model to first reason about which sections are most relevant to the given question, and then synthesize an answer based on the retrieved segments. This combination of structured retrieval and reasoning-aware prompting may help mitigate the tendency toward retrieval preference.

\subsection{Generality of LLMs}
Current LLMs are primarily trained as general-purpose systems. As a result, they may lack sufficient domain-specific training data to produce highly specialized or nuanced reviews in certain academic fields. A potential solution to improve the use of LLMs in paper review is to develop fine-tuned domain-specific models. For a particular conference or journal, historical submission data—including both accepted and rejected papers—could be used to fine-tune the LLM. The resulting specialized model may be better suited to understand the standards and expectations of the target venue, thereby improving its ability to assist in the review process more accurately and reliably.
\section{Conclusion}
\label{sec:conclusion}
This work explores the feasibility of employing LLMs in the academic paper review process. It proposes a system for LLM-based reviewing, implements a prototype system, and releases the corresponding code as open source. Experiments conducted on submissions to the WASA conference demonstrate that the overlap between papers accepted by the LLM-based system and those accepted by human reviewers averages only 38.6\%. Further analysis reveals critical issues in LLM-based reviewing, including a lack of independent judgment and a tendency toward retrieval preferences. Given these limitations, we recommend that LLMs be used as assistive tools to support human reviewers, enhancing efficiency while leaving the final decision-making to domain experts.

\section{Ethical Consideration}
All WASA papers used in this work were accessed with the explicit approval of the conference organizers. Throughout the research process, we strictly adhered to ethical guidelines regarding the use of unpublished academic papers. No ideas, methods, or intellectual contributions from the authors were misappropriated or presented as our own. Moreover, no rejected or unpublished submissions were redistributed or shared externally in any form. The use of this data was solely for experimental and analytical purposes, and every effort was made to respect author confidentiality and intellectual property. 



\bibliographystyle{IEEEtran}
\bibliography{references}

\end{document}